\documentclass[12pt]{article}
\usepackage{fullpage,authblk,cite}
\usepackage{amsmath,amsthm,graphicx,amssymb,verbatim,listings}
\usepackage{wasysym,color,enumitem,mathcomp}
\usepackage[T1]{fontenc}     
\usepackage{lmodern, tipa}   
\usepackage[ pdftex, plainpages = false, pdfpagelabels,
                 pdfpagelayout = useoutlines,
                 bookmarks,
                 bookmarksopen = true,
                 bookmarksnumbered = true,
                 breaklinks = true,
                 linktocpage=all,
                 pagebackref=false,
                 colorlinks = true,
                 linkcolor = Blue,
                 urlcolor  = blue,
                 citecolor = BrickRed,
                 anchorcolor = green,
                 hyperindex = true,
                 hyperfigures
                 ]{hyperref}
\usepackage[usenames, dvipsnames]{xcolor}
\usepackage{xifthen} 

\newcommand{\tr}{\hbox{tr}}

\newcommand{\ket}[1]{{\ensuremath{\left| #1 \right\rangle}}}
\newcommand{\bra}[1]{{\ensuremath{\left\langle #1 \right|}}}

\newcommand{\ketbra}[2]{{\ensuremath{\left| #1 \middle\rangle \middle\langle #2
      \right|}}}

\newcommand{\arxiv}[2][]{\ifthenelse{\isempty{#1}}{\href{http://arxiv.org/abs/#2}{{\tt arXiv:\allowbreak{}#2}}} {\href{http://arxiv.org/abs/#2}{{\tt arXiv:\allowbreak{}#2 [#1]}}}}
\newcommand{\pirsa}[1]{\href{http://pirsa.org/#1/}{{\tt PIRSA:\allowbreak{}#1}}}
\newcommand{\booktitle}{\textsl}
\newcommand{\hrefdoi}[2]{\href{https://dx.doi.org/#1}{#2}}

\newcommand{\bbR}{\mathbb{R}}
\newcommand{\bbC}{\mathbb{C}}

\begin{document}
\title{Further Exercises about Sporadic SICs}
\author[$\dag$]{Blake C.\ Stacey}
\affil[$\dag$]{Physics Department, University of
    Massachusetts Boston\protect\\ 100 Morrissey Boulevard, Boston MA 02125, USA}

\date{\small\today}

\maketitle

\begin{abstract}
I review some recent technical developments in quantum information
theory by rephrasing them in the form of exercises.
\end{abstract}

A couple years ago, I officially became a textbook
author~\cite{Stacey:2021}. For reasons that made sense at the time, I
gathered all the homework problems together at the end into a chapter
called, well, ``Exercises''. And now I keep getting spam invitations
to conferences and special issues of journals no one has ever heard
of, asking me to share my pivotal work ``in the field of Exercises''.

The book grew out of an effort to find a way to present results that
did not fit neatly into the journal-article form factor (e.g.,
\cite{Stacey:2016}). Because these results reside at the intersection
of multiple subject areas, setting the context appropriately meant
explaining multiple background topics, at least in outline, and so the
natural length of the work tended to the monograph scale. I tried to
provide a mix of homework problems ranging from fairly direct
calculations to more open-ended puzzles. Having spent more time both
doing research and teaching since then, I have devised further
problems that likewise prod outward in different directions from the text.

The subject area of the book is the topic of \emph{Symmetric
  Informationally Complete quantum measurements,} known for short as
\emph{SICs.} In geometrical terms, a SIC is a set of $d^2$ unit
vectors in a $d$-dimensional complex vector space with the property
that the absolute value of the inner product between any pair is the
same as that for any other pair~\cite{Renes:2004, Scott:2010,
  Fuchs:2017}. The \emph{sporadic} SICs are the examples in dimensions
2 and 3, along with some of the examples in dimension 8, which stand
out in various ways from the other known
SICs~\cite{Appleby:2017b}. Unfortunately, the general question of SIC
existence in arbitrary dimension is not yet sufficiently understood
that it can be made into homework problems~\cite{Horodecki:2022}.

\vfill\eject

\noindent\textbf{Problem 1: Nonexistence of Triply Transitive SICs}

The qubit, Hesse and Hoggar SICs are \emph{doubly transitive}: Any
pair of rays can be mapped into any other by some symmetry of the
SIC. This naturally raises the question of whether any \emph{triply}
transitive SICs exist.

Define the \emph{triple products} of a SIC $\{\Pi_j\}$ to be the
quantities
\begin{equation}
  T_{jkl} := \tr(\Pi_j \Pi_k \Pi_l) \, .
\end{equation}

\begin{itemize}
  \item[(a)] Evaluate $T_{jkl}$ in the case that all three indices are
    equal and in the case that two out of three indices are
    equal. Find the magnitude $|T_{jkl}|$ in the case that all three
    indices are different.

  \item[(b)] Using your results from part (a), evaluate the sum
    $\sum_l T_{jkl}$, assuming that $j \neq k$.

  \item[(c)] Show that the normalized triple products
    \begin{equation}
      \tilde{T}_{jkl} := \frac{T_{jkl}}{|T_{jkl}|}
    \end{equation}
    satisfy the relation
    \begin{equation}
      \tilde{T}_{jkl} = \tilde{T}_{mjk} \tilde{T}_{mkl}
      \tilde{T}_{mlj} \, .
    \end{equation}
    
  \item[(d)] Now suppose that the SIC $\{\Pi_j\}$ is triply
    covariant. Show that the normalized triple products for distinct
    indices $j,k,l$ are all equal, and find the values that they can
    possibly take. Use this to evaluate the sum $\sum_l T_{jkl}$, and
    show that it is inconsistent with the answer to part~(b).
\end{itemize}

This problem is based on a paper by Zhu~\cite{Zhu:2015}.

\medskip

\noindent\textbf{Problem 2: Entanglement Properties of the Hoggar
  Lines}

\begin{itemize}
\item[(a)] A Hoggar-type SIC is constructed as the orbit of a 
  vector $\ket{\pi_0} \in \bbC^8$ under the action of the three-qubit
  Pauli group. Convince yourself that this implies that all the
  vectors in the SIC are equally entangled.

\item[(b)] Evaluate the 3-tangle~\cite{Coffman:2000} of the Hoggar
  fiducial, Eq.~(2.37).

\item[(c)] The \emph{Fano representation} of an $n$-qubit state is
  found by using the $n$-qubit Pauli operators
  \begin{equation}
    \sigma_\alpha := \sigma_{\alpha_1} \otimes \sigma_{\alpha_2}
    \otimes \cdots \otimes \sigma_{\alpha_n}
  \end{equation}
  as an operator basis:
  \begin{equation}
    \rho = \frac{1}{2^n} \sum_\alpha c_\alpha \sigma_\alpha \, ,
  \end{equation}
  where $\sigma_0$ is understood to be the $2\times2$ identity
  matrix. What are the absolute values of the coefficients
  $\{c_\alpha\}$ in the Fano representation of a Hoggar state?

\item[(d)] What happens to the Fano representation of a 3-qubit state
  when we take the partial trace over one of the qubits? What does
  this imply for the partial trace of the Hoggar fiducial projector
  over any of the three qubits?

\item[(e)] What happens if we take the partial trace of the Hoggar
  fiducial projector over two of the three qubits? Where do the
  resulting single-qubit states live in the Bloch ball picture?
  
\end{itemize}

\medskip

\noindent\textbf{Problem 3: Unbiased Rank-1 Qubit MICs}

An unbiased rank-1 qubit MIC is a set of four operators $\{E_0, E_1,
E_2, E_3\}$ that sum to the identity
\begin{equation}
  \sum_j E_j = I
\end{equation}
and which are each proportional to a projection onto a vector:
\begin{equation}
  E_j = \frac{1}{2} \ketbra{\psi_j}{\psi_j} \, .
\end{equation}
Each projection operator $\ketbra{\psi_j}{\psi_j}$ has a
representation as a point on the Bloch sphere. Let $\vec{r}_j$ be the
vector from the origin to the point representing
$\ketbra{\psi_j}{\psi_j}$.

\begin{itemize}
\item[(a)] Show that
  \begin{equation}
    \sum_j \vec{r}_j = 0 \, .
  \end{equation}
  What does this imply for the dot products of disjoint pairs, e.g.,
  $\vec{r}_0 \cdot \vec{r}_1$ and $\vec{r}_2 \cdot \vec{r}_3$?

\item[(b)] The four vectors $\{\vec{r}_j\}$ are the vertices of a
  tetrahedron inscribed in the Bloch sphere. Find the center point of
  each face of this tetrahedron. How far are these from the origin?

\item[(c)] Verify that for any qubit state $\rho$, at most one of the
  $\{E_j\}$ can satisfy $\tr \rho E_j = 0$. From this, prove that the
  probability vectors that represent any two states $\rho$ and $\rho'$
  can never be orthogonal.
\end{itemize}

This problem arose from~\cite{DeBrota:2019} and \cite{Fuchs:2023}.

\medskip

\noindent\textbf{Problem 4: Orthocross MICs}

The \emph{orthocross} MICs are constructed by starting with an
orthonormal basis $\{\ket{j}\}$ for $\bbC^d$, building a set of
``cross terms'' and squishing the result so that it forms a
POVM~\cite{DeBrota:2021}. They were first introduced in order to prove
the quantum de Finetti theorem~\cite{Caves:2002}. The construction
proceeds as follows. First, we take the $d$ projectors
$\{\ketbra{j}{j}\}$, and then we find all the products
\begin{equation}
\frac{1}{2}\left(\ket{j} + \ket{k}\right)
 \left(\bra{j} + \bra{k}\right)
\end{equation}
where $j < k$. We next construct all the products of the form
\begin{equation}
\frac{1}{2}\left(\ket{j} + i\ket{k}\right)
 \left(\bra{j} - i\bra{k}\right) \, ,
\end{equation}
where again the indices satisfy $j < k$. The union of these three sets
of projectors we'll call $\{\Pi_\alpha\}$. Together, $\{\Pi_\alpha\}$
forms a positive semidefinite operator basis for the space of
self-adjoint operators on $\bbC^d$. To make this operator basis into a
POVM, we evaluate
\begin{equation}
S = \sum_{\alpha=1}^{d^2} \Pi_\alpha \, ,
\end{equation}
and then take
\begin{equation}
    E_\alpha := S^{-1/2} \Pi_\alpha S^{-1/2} \, .
\end{equation}

For $d = 2,3$ and 8, construct an orthocross MIC $\{E_\alpha\}$ and
evaluate its Born matrix $\Phi$. What is $||I - \Phi||_2^2$, and how
does it compare with $||I - \Phi_{\mathrm{SIC}}||_2^2$?

\medskip

\noindent\textbf{Problem 5: Qubit Clifford Group}

Let $\sigma_x$, $\sigma_y$ and $\sigma_z$ be the Pauli matrices, and
define the \emph{phase} and \emph{Hadamard} unitary matrices by
\begin{equation}
  S := \begin{pmatrix} 1 & 0 \\ 0 & i \end{pmatrix} \, ,
  \ H := \frac{1}{\sqrt{2}} \begin{pmatrix} 1 & 1 \\ 1 & -1
  \end{pmatrix} \, .
\end{equation}

\begin{itemize}
  \item[(a)] Find the result of conjugating each of the Pauli matrices
    by $S$ and by $H$, i.e., evaluate $H \sigma_j H^\dag$ and $S
    \sigma_j S^\dag$ for $j = x,y,z$.

  \item[(b)] Show that the set of all matrix products $MM'$, where
    \begin{equation}
      M \in \{I, H, S, HS, SH, HSH\}
    \end{equation}
    and
    \begin{equation}
      M' \in \{I, \sigma_x, \sigma_y, \sigma_z\} \, ,
    \end{equation}
    forms a group of order 24. This is the Clifford group for a single
    qubit.

  \item[(c)] What is the orbit of the vector
    \begin{equation}
      \ket{z+} := \begin{pmatrix} 1 \\ 0 \end{pmatrix}
    \end{equation}
    under the action of this group? That is, what set of states is
    constructed by taking the products $U\ket{z+}$ for all the $U$ in
    the group?

  \item[(d)] What is the orbit of the SIC fiducial
    \begin{equation}
      \Pi_0 := \frac{1}{2}\left(I + \frac{1}{\sqrt{3}}
      (\sigma_x + \sigma_y + \sigma_z)\right)
    \end{equation}
    under the action of this group? That is, what set of quantum
    states do we get by taking $U\Pi_0U^\dag$ for all the $U$ in the
    group?
    
\end{itemize}

\medskip

\noindent\textbf{Problem 6: More on Conjugate Qubit SICs}

Let $\{\Pi_j : j = 0,1,2,3\}$ be an arbitrary qubit SIC. Then any
qubit density matrix $\rho$ can be written as
\begin{equation}
  \rho = \sum_j \left[3p(j) - \frac{1}{2}\right]\Pi_j \, ,
\end{equation}
where
\begin{equation}
  p(j) := \frac{1}{2} \tr(\rho\Pi_j) \, .
\end{equation}

\begin{itemize}
  \item[(a)] Show that the four operators given by
    \begin{equation}
      \Pi_j' := I - \Pi_j
    \end{equation}
    also form a SIC.

  \item[(b)] Find the probabilities for the outcomes of the new SIC
    measurement in terms of those for the old. Show that there is no
    way to write this transformation as merely a \emph{rotation} of
    the state space, i.e., without doing a reflection.

  \item[(c)] Using the specific SIC defined in Eq.~(2.31), find what
    the transformation in part (b) does to the probabilities in
    Eq.~(2.32). How are complex conjugation, Galois conjugation and
    the Clifford group all related here?
    
\end{itemize}

\medskip

\noindent\textbf{Problem 7: Qutrit Clifford Transformations}

Let $X$ and $Z$ be the qutrit Weyl--Heisenberg operators:
\begin{equation}
  X\ket{j} = \ket{j+1 \mod 3},
  \ Z\ket{j} = \omega^j \ket{j} \, ,
\end{equation}
with $\omega := e^{2\pi i/3}$. Consider the unitary matrix
\begin{equation}
  U := \frac{1}{\sqrt{3}} \begin{pmatrix}
    1 & 1 & 1 \\
    1 & \omega & \omega^2 \\
    1 & \omega^2 & \omega
  \end{pmatrix} \, .
\end{equation}

\begin{itemize}
  \item[(a)] Show that multiplying $U$ into a vector implements a
    discrete Fourier transform of that vector.

  \item[(b)] Find the result of conjugating $X$ and $Z$ by $U$, i.e.,
    evaluate $U^\dag X U$ and $U^\dag Z U$. Prove that conjugating any
    product of the form $X^a Z^b$ by $U$ will, up to an overall phase,
    result in a product of the same form.

  \item[(c)] Show that the Hesse SIC fiducial vector is an
    eigenvector of $U$, and find its eigenvalue.

  \item[(d)] Repeat your analysis in parts (b) and (c) using the
    \emph{Zauner} unitary
    \begin{equation}
      V := \frac{e^{i\pi/6}}{\sqrt{3}}
      \begin{pmatrix}
        1 &  1 &  1 \\
        \omega^2 & 1 & \omega \\
        \omega^2 & \omega & 1
      \end{pmatrix} \, .
    \end{equation}
\end{itemize}

\medskip

\noindent\textbf{Problem 8: Variant QBic Equations}

Because a SIC furnishes a basis for operator space, we can write a
product of two SIC projectors as a sum:
\begin{equation}
  \Pi_j \Pi_k = \sum_{l} \alpha_{jkl} \Pi_l \, ,
\end{equation}
where the $\{\alpha_{jkl}\}$ are the \emph{structure coefficients} of
the basis.

\begin{itemize}
  \item[(a)] Find the relation between the structure coefficients and
    the triple products.

  \item[(b)] Show that the QBic equation (2.20), which derives from
    the condition $\tr\rho^3 = 1$, is equivalent to
    \begin{equation}
      \sum_{jkl} (\mathrm{Re}\, \alpha_{jkl}) p(j) p(k) p(l)
      = \frac{4}{d(d+1)^2} \, .
      \label{eq:qbic-alternate}
    \end{equation}

  \item[(c)] Take $d = 3$, and let $\{\Pi_j\}$ be an arbitrary qubit
    SIC (i.e., not necessarily the Hesse SIC). Show that
    Eq.~(\ref{eq:qbic-alternate}) reduces to
    \begin{equation}
      \frac{1}{2} \sum_j p(j)^3 + \sum_{j \neq k \neq l}
      (\mathrm{Re}\, \alpha_{jkl}) p(j) p(k) p(l) = 0 \, .
    \end{equation}
\end{itemize}
This exercise was inspired by Tabia and Appleby~\cite{Tabia:2013}.

\medskip

\noindent\textbf{Problem 9: Relating One and Three Qubits}

Let $\ket{\pi_+}$ be the eigenvector of the qubit SIC fiducial
projector $\Pi_0$ used in Problem~6(c), and let $\ket{\pi_-}$ be a
vector orthogonal to it. Show that
\begin{equation}
  \ket{\pi} = a \ket{\pi_+}\otimes\ket{\pi_+}\otimes\ket{\pi_+}
  + b\ket{\pi_-}\otimes\ket{\pi_-}\otimes\ket{\pi_-} \, ,
\end{equation}
where $a,b \in \bbR$, is a Hoggar-type fiducial when $a^2 + b^2 = 1$
and $a^2 - b^2 = \frac{1}{\sqrt{3}}$.

This problem is based on a talk by Wootters~\cite{Wootters:2009}.

\medskip

\noindent\textbf{Problem 10: A Symmetric Reference Measurement of
  Higher Rank}

Suppose that the dimension $d$ is odd. The Weyl--Heisenberg
displacement operators are
\begin{equation}
  D_{a,b} = (-e^{i\pi/d}) X^a Z^b \, ,
\end{equation}
where we have chosen the phase prefactor for convenience's
sake. Consider the operator found by summing over all of these except
the identity:
\begin{equation}
  B = \frac{1}{\sqrt{d+1}} \sum_{a,b \neq 0,0} D_{a,b} \, ,
\end{equation}
and find all of its Weyl--Heisenberg conjugates:
\begin{equation}
B_{a,b} = D_{a,b} B D_{a,b}^\dag \, .
\end{equation}
Define a set of operators $\{E_{a,b}\}$ by
\begin{equation}
  E_{a,b} = \frac{1}{d^2}
  \left(I + \frac{1}{\sqrt{d+1}} B_{a,b}\right) \, .
\end{equation}

\begin{itemize}
  \item[(a)] Show that $\{E_{a,b}\}$ is an informationally complete
    POVM where each element has rank $(d+1)/2$.

  \item[(b)] Take $d = 3$. How does the POVM constructed here compare
    with the Hesse SIC?
\end{itemize}

This exercise was based on work relating POVMs and Wigner
functions~\cite{Appleby:2007,DeBrota:2020}.

\medskip

\noindent\textbf{Problem 11: More on Conjugate Hoggar-type SICs}

Eqs.~(2.37) and (5.49) give fiducial vectors that are the same apart
from a complex conjugation and that both generate Hoggar-type
SICs. Let $\Pi_0^+$ be the projector onto the fiducial with the $+$
choice of sign and $\Pi_0^-$ be the projector onto the fiducial with
the $-$ choice.

\begin{itemize}
  \item[(a)] What is the relation between the Fano representations of
    $\Pi_0^+$ and $\Pi_0^-$?

  \item[(b)] Consider the fiducial projector $\Pi_0^+$. Suppose we
    change the signs on all the nontrivial terms of its Fano
    representation, i.e., all the terms other than $I \otimes I
    \otimes I$. What kind of matrix does this yield?

  \item[(c)] What kind of structure results from taking the orbit
    under the three-qubit Pauli group of the matrix found in
    part~(b)?
\end{itemize}

\medskip

\noindent\textbf{Problem 12: Nonstandard State Updates}

In chapter 2, we discussed how to understand the quantum theory of a
$d$-dimensional system using a \emph{reference measurement} with $d^2$
outcomes~\cite{Appleby:2017, DeBrota:2021b}. This leads to a
probabilistic representation of the Born rule, Eq.~(2.17):
\begin{equation}
  q(D) = p(D|R) \, \Phi \, p(R) \, ,
\end{equation}
where the Born matrix $\Phi$ is defined by the reference POVM
$\{R_i\}$ and its corresponding post-measurement states
$\{\sigma_i\}$:
\begin{equation}
  [\Phi^{-1}]_{ij} := \tr(R_i \sigma_j) \, .
\end{equation}
When the reference measurement $\{R_i\}$ is a SIC and the
$\{\sigma_i\}$ are the projectors proportional to the POVM elements
(\emph{parallel updating}), the Born rule becomes the familiar
expression
\begin{equation}
  q(D_j) = \sum_i\left[(d+1)p(R_i) - \frac{1}{d}\right] p(D_j|R_i) \,
  .
\end{equation}
\begin{itemize}
\item[(a)] Take $d = 2$ and let $\{R_i\}$ be the qubit SIC generated
  from the fiducial $\Pi_0^+$ (as in Problem~11). For the
  $\{\sigma_i\}$, take the states orthogonal to those:
  \begin{equation}
    \sigma_i = I - \frac{R_i}{\tr R_i} \, .
    \label{eq:antipodal}
  \end{equation}
  What is the Born matrix $\Phi$? What is the squared Frobenius norm
  $||I-\Phi||_2^2$ in this case?

\item[(b)] Suppose now that the post-measurement states $\{\sigma_i\}$
  are a SIC produced by a \emph{random} unitary conjugation of the
  original. How does $||I-\Phi||_2^2$ behave?

\item[(c)] What if $\{\sigma_i\}$ is produced by applying a random
  unitary to the SIC constructed from the orthogonal complements of
  the original, as in Eq.~(\ref{eq:antipodal})?
\end{itemize}
Parts (b) and (c) are amenable to numerical investigation. This
problem was inspired by~\cite{DeBrota:2020b}, and in particular the
lessons that you should gather numerical evidence for a conjecture
before you try to prove it, as well as after you believe you have
proved it.

\bigskip

\noindent\textbf{Acknowledgments}

This work was supported by National Science Foundation Grant 2210495.

\end{document}